\documentclass[twocolumn,amsmath,pra]{revtex4-1}
\usepackage{amsthm,hyperref,cleveref,graphicx}
\newcommand{\Pp}{\mathcal{P}}
\newcommand{\Ms}{\mathcal{M}}
\newcommand{\PI}{\Pp_i}
\newcommand{\MJ}{\Ms_j}
\newtheorem{thm}{Theorem}
\newtheorem{lem}{Lemma}
\DeclareMathOperator{\tr}{Tr}
\begin{document}

\title{Robust preparation noncontextuality inequalities in the simplest scenario}
\author{Matthew F. Pusey}
\affiliation{Department of Computer Science, University of Oxford,  Wolfson Building,  Parks Road, Oxford OX1 3QD, UK}
\affiliation{Perimeter Institute for Theoretical Physics, 31 Caroline Street North, Waterloo, ON N2L 2Y5, Canada}
\date{Aug 2, 2018}
\begin{abstract}
  Contextuality is the leading notion of nonclassicality for a single system. However, an experimental demonstration requires finding procedures that are operationally equivalent, which might seem impossible to achieve exactly. Here I focus on the simplest non-trivial case, four preparations and two tomographically complete binary measurements. Exploiting a subtle connection to the CHSH scenario gives eight non-linear inequalities which are together necessary and sufficient for the experimental statistics to admit a preparation noncontextual model in such a scenario. No fixed operational equivalences are required, removing a key difficulty with experimental tests of older preparation noncontextuality inequalities.
\end{abstract}
\maketitle

\section{Introduction}
The gold standard for an experiment that defies classical explanation is the violation of a Bell inequality \cite{bell,bellreview}. In the case of a single quantum system, this is not a possibility, and so attention has focussed on contextuality.

Contextuality was first identified by Bell, Kochen and Specker \cite{bellks,ks}. Whilst this was a profound insight into quantum mechanics, the definition they used is stated in quantum terms, and applies only to the ideal of projective measurements. It is therefore not amenable to experimental test. A generalised definition due to Spekkens \cite{cntx} is stated operationally and applies to arbitrary procedures. As shown in \cite{parity}, this definition, or even just one component of it known as \emph{preparation noncontextuality}, is thus well suited to experiment. (For an alternative perspective on experimental contextuality, not based on Spekkens' generalisation, see for example \cite{klyacho08,kirchmair09,guhne10,lapkiewicz11,mod1,mod2}.)

However, \cite{parity} used an assumption that two preparation procedures were indistinguishable, which was not satisfied exactly in the reported experiment and never will be in any experiment. Here I show how this problem can be eliminated by providing a full characterisation of the preparation noncontextual statistics in the simplest scenario to which the concept applies. The shift in approach is that, with the help of tomographically complete measurements, indistinguishable preparation procedures are inferred from the statistics, rather than posited \emph{a priori}.

\section{Definitions}\label{definitions}
Consider an experiment where one implements a preparation procedure $\PI$ followed by a measurement procedure $\MJ$ with outcome $k$, characterised by the probabilities $P(k | \PI, \MJ)$. An \emph{ontological model} seeks to explain these results via an \emph{ontic state} $\lambda$ that screens off the preparation from the measurement result:
\begin{equation}
  P(k | \PI, \MJ) = \int P(k | \lambda, \MJ) \mu_i (\lambda)d\lambda,
  \label{onticop}
\end{equation}
where we use the shorthand $\mu_i(\lambda) = P(\lambda | \PI)$.

The explanation proffered by an ontological model is compelling only if it does justice to important features of the observed statistics. For example, in a bipartite scenario, Bell's locally causal models would provide a natural explanation for the observed no-signalling \cite{tuning}.

Preparation noncontextuality concerns a closely related feature, namely \emph{operational equivalence} among preparations. Of particular relevance are operationally equivalent mixtures: suppose there exists probability distributions $\{p_i\}$ and $\{q_i\}$ such that for all $j, k$
\begin{equation} \sum_i p_i P(k|\PI, \MJ) = \sum_i q_i P(k|\PI, \MJ). \label{opequiv}
\end{equation}
Then we say that the probabilistic mixtures, which might be written $\sum_i p_i \PI$  and $\sum_i q_i \PI$, are operationally equivalent. In principle it is possible that somebody invents a new measurement procedure $\Ms'$ such that $\sum_i p_i P(k | \PI, \Ms') \neq \sum_i q_i P(k|\PI, \Ms')$, in which case the apparent operational equivalence would evaporate. For the time being we will assume no such $\Ms'$ exists. In other words, we assume that the $\MJ$ are \emph{tomographically complete} or \emph{fiducial} \cite{lucien} for the $\PI$.

This assumption can be made without specifying any particular operational theory, but as an example: in the language of quantum theory, where preparations $\PI$ are associated with density operators $\rho_i$, we need that $\sum_i p_i \rho_i = \sum_i q_i \rho_i$. This follows from \cref{opequiv} if and only if the POVM elements associated with the $\MJ$ span the space of operators defined by the $\rho_i$ \cite{tomocomplete}.

  What can explain the inability of any measurement to distinguish $\sum_i p_i \PI$ from $\sum_i q_i \PI$? The most natural explanation is that this ``distinction without a difference'' is no distinction at all:
\begin{equation}
  \sum_i p_i \mu_i(\lambda) = \sum_i q_i \mu_i(\lambda).\label{onticequiv}
\end{equation}
The inference from the operational equivalence \eqref{opequiv} to the ontic equivalence \eqref{onticequiv} constitutes the assumption of preparation noncontextuality \cite{cntx}. (For comparison, measurement noncontextuality is the assumption that measurements that cannot be distinguished by the statistics for any preparation are equivalent in the ontological model.) Note that this assumption is presented slightly differently in \cite{cntx,parity}, for readers familiar with the latter presentation the connection is made in \cref{formulating}.

This article will focus on four preparations and two binary measurements. This is the simplest non-trivial scenario because, as shown in \cref{simplest}, there is a noncontextual model for \emph{any} operational probabilities in any simpler scenario. In a scenario with two binary measurements the operational probabilities for a single preparation $\PI$ are given by the 2-dimensional real vector $\vec \PI = (P(0|\PI,\Ms_0)-P(1|\PI,\Ms_0), P(0|\PI,\Ms_1)-P(1|\PI,\Ms_1))$ which (along with normalisation) fixes all 4 probabilities. 

\begin{figure}
  \includegraphics[scale=0.8]{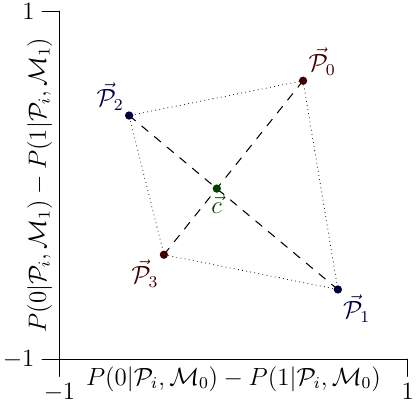}
  \caption{Example statistics. The preparations have been labelled in accordance with the conventions that $\vec \Pp_0$ is opposite $\vec \Pp_3$ and $\{\vec \Pp_0 - \vec \Pp_3, \vec \Pp_2 - \vec \Pp_1\}$ is positively oriented. Also shown is $\vec c$ as defined by \cref{pivotalop}.}
  \label{conventions}
\end{figure}

\section{Labelling conventions}
Denoting the four preparations $\{\Pp_0, \Pp_1, \Pp_2, \Pp_3\}$, the $\vec \PI = (x_i, y_i)$ must be the vertices of a convex quadrilateral, since any degeneracy will lead to a simplex and hence an immediate preparation noncontextual model by the argument in \cref{simplest}. As in \cref{conventions}, we adopt the conventions that $\vec \Pp_0$ is opposite to $\vec \Pp_3$, and the $\vec \Pp_0 - \vec \Pp_3$ and $\vec \Pp_2 - \vec \Pp_1$ diagonals are positively oriented:
\begin{equation}
  \begin{vmatrix}
    x_0 - x_3 & x_2 - x_1 \\
    y_0 - y_3 & y_2 - y_1
  \end{vmatrix} > 0.
  \label{orientation}
\end{equation}

\section{A pivotal equivalence}\label{pivotalsec}
One example of an operational equivalence is given by the point $\vec c$ at which the $\{\vec \Pp_0, \vec \Pp_3\}$ diagonal intersects $\{\vec \Pp_1, \vec \Pp_2\}$ diagonal, giving probabilities $p, q$ such that 
\begin{equation}
p\vec \Pp_0 + (1-p)\vec \Pp_3 = q\vec \Pp_1 + (1-q)\vec \Pp_2 = \vec c. \label{pivotalop}
\end{equation} Preparation noncontextuality then demands that
\begin{equation}
  p\mu_0(\lambda) + (1-p)\mu_3(\lambda) = q\mu_1(\lambda) + (1-q)\mu_2(\lambda) \label{pivotal}.
\end{equation}
  I will now show that in the current scenario, this single equivalence is in fact sufficient for a model to be preparation noncontextual.

  Suppose we have a preparation contextual model, i.e. there exists $p_i, q_i$ such that \cref{opequiv} holds yet \cref{onticequiv} fails. We want to prove that \cref{pivotal} must also fail. The first step is to show that \cref{onticequiv} must fail for some $p'_i, q'_i$ with $\sum_i p'_i \vec \PI = \sum_i q'_i \vec \PI = \vec c$. To see this, denote $\vec p = \sum_i p_i \vec \PI = \sum_i q_i \vec \PI$, and notice that since $\vec c$ is in the interior of the quadrilateral $\vec c = \sum_i r_i \vec \PI + r_* \vec p$ for some probability distribution $\{r_0, r_1, r_2, r_3, r_*\}$ with $r_* > 0$. But then $p_i' = r_i + r_* p_i$ and $q_i' = r_i + r_* q_i$ give the required instance, with the failure of \eqref{onticequiv} ensured by $r_* > 0$ and the fact that \eqref{onticequiv} fails for the $p_i, q_i$.

  Now I will argue that there exist probabilities $s, t$ such that $\sum_i p'_i \PI$ amounts to preparing $p \Pp_0 + (1-p) \Pp_3$ with probability $s$ and $q \Pp_1 + (1-q) \Pp_2$ with probability $(1-s)$ (similarly for the $q'_i$ with $t$ in place of $s$). Formally, this means $\{p'_0, p'_1, p'_2, p'_3\} = \{s p, (1-s)q, (1-s)(1-q), s(1-p)\}$ and $\{q'_0, q'_1, q'_2, q'_3\} = \{t p, (1-t)q, (1-t)(1-q), t(1-p)\}$.  If that is the case then \eqref{pivotal} implies $\sum_i p'_i \mu_i(\lambda) = \sum_i q'_i \mu_i(\lambda)$ and so the failure of the latter requires the failure of the former, which is what we wanted to prove.

  To find $s$ and $t$ it is useful to make an affine transformation to a new co-ordinate system in which $\vec c = (0,0)$, $\vec \Pp_0 = (1-p,0)$ and $\vec \Pp_1 = (0,1-q)$. Then $p\vec \Pp_0 + (1-p)\vec \Pp_3 = \vec c$ gives $\vec \Pp_3 = (-p,0)$ and $q\vec \Pp_1 + (1-q)\vec \Pp_2 = \vec c$ gives $\vec \Pp_2 = (0, -q)$. Now $\sum_i p'_i \vec\PI = \vec c$ becomes $(p'_0 (1-p) - p'_3 p, p'_1 (1-q) - p'_2 q) = (0,0)$. Defining $s = p'_0/p \geq 0$ we have $p'_3 = s(1-p)$. Similarly defining $\bar s = p'_1/q \geq 0$ we have $p'_2 = \bar s(1-q)$. $\sum_i p'_i = 1$ gives $s + \bar s = 1$, and repeating the same argument with the $q'_i$ gives $t$.

\section{The connection to CHSH}\label{chshsec}
It was first shown by Barrett that the existence of a preparation noncontextual model for a single system implies the existence of a locally causal model for any bipartite scenario involving that system \cite{jon}, in other words, any bipartite proof of Bell's theorem is a proof of preparation contextuality. The converse is not expected to hold in general (although certain proofs of preparation contextuality can be converted into bipartite proofs of Bell's theorem \cite{matt}). Nevertheless, with the reduction of the previous section in hand, Barrett's argument can be extended to see that the existence of a preparation noncontextual model for four preparations and two tomographically complete binary measurements is \emph{equivalent} to the existence of a Bell local model in the scenario considered by CHSH \cite{chsh}.

In the relevant Bell scenario two parties choose between two measurements, their choices labelled $x$ and $y$. They both obtain a binary outcome, labelled $a$ and $b$.  Their statistics $P(a,b|x,y)$ are related to the preparation noncontextuality scenario by
\begin{multline}
  \begin{pmatrix}
    P(0,k|0,j) & P(1,k|0,j) \\
    P(0,k|1,j) & P(1,k|1,j) \\
  \end{pmatrix} \\= 
  \begin{pmatrix}
    p P(k|\Pp_0, \MJ) & (1-p) P(k|\Pp_3, \MJ) \\
    q P(k|\Pp_1, \MJ) & (1-q) P(k|\Pp_2, \MJ)
  \end{pmatrix}.
  \label{opconversion}
\end{multline}

These statistics are normalised, and are no-signalling due to the operational equivalence \eqref{pivotalop}. If we have a preparation noncontextual model, then set $\mu(\lambda) = p\mu_0(\lambda) + (1-p)\mu_3(\lambda)$,
\begin{multline}
  \begin{pmatrix}
    P_A(0|\lambda,0) & P_A(1|\lambda,0) \\
    P_A(0|\lambda,1) & P_A(1|\lambda,1) \\
  \end{pmatrix} \\= 
  \begin{pmatrix}
    p\mu_0(\lambda)/\mu(\lambda) & (1-p)\mu_3(\lambda)/\mu(\lambda)  \\
    q\mu_1(\lambda)/\mu(\lambda) & (1-q)\mu_2(\lambda)/\mu(\lambda)
  \end{pmatrix}
  \label{conversion}
\end{multline}
(which is normalised by \cref{pivotal}), and $P_B(k|\lambda,j) = P(k|\lambda,\MJ)$. Then \cref{onticop} gives the locally causal model
\begin{equation}
  p(a,b|x,y) = \int P_A(a|\lambda,x)P_B(b|\lambda,y)\mu(\lambda)d\lambda.
\end{equation}

If, conversely, we start with a locally causal model, then inverting \cref{conversion} gives an ontological model for $P(k|\PI,\MJ)$, where \cref{pivotal} is guaranteed by the normalisation of the locally causal model. Since we have seen that \cref{pivotal} is sufficient for a preparation noncontextual model, we have established the desired equivalence.

Fine \cite{fine} has shown that the eight versions of the CHSH inequality \cite{chsh} are necessary and sufficient for the existence of a locally causal model. Hence given $P(k|\PI,\MJ)$, one can calculate the corresponding Bell scenario probabilities \eqref{opconversion} and then use the eight CHSH inequalities to determine whether or not a preparation noncontextual model exists.

\section{A closed expression}\label{closedsec}
The above argument completely characterises the noncontextual statistics in our scenario. However, it might appear that if one were to calculate $p$ and $q$ explicitly and substitute \eqref{opconversion} into the CHSH inequalities the result would be extremely convoluted. In fact it can be written in a remarkably simple form, thanks to following lemma.

\begin{lem}
Suppose the $\vec \PI = (x_i, y_i)$ for $i={0,1,2,3}$ satisfy \cref{orientation}, and $p$ and $q$ are defined as the solutions of \cref{pivotalop}. Then for any real numbers $\{z_0,z_1,z_2,z_3\}$,
\begin{equation}
  pz_0 + (1-p)z_3 \leq qz_1 + (1-q)z_2
  \label{lem1}
\end{equation}
if and only if
\begin{equation}
\begin{vmatrix}
x_0 & y_0 & z_0 & 1 \\
x_1 & y_1 & z_1 & 1 \\
x_2 & y_2 & z_2 & 1 \\
x_3 & y_3 & z_3 & 1
\end{vmatrix} \leq 0.
\label{lem2}
\end{equation}
Furthermore, equality in \eqref{lem1} and \eqref{lem2} is also equivalent.
\end{lem}
Geometrically this lemma concerns a tetrahedron with vertices $(x_i, y_i, z_i)$. \Cref{lem1} asks whether, when the 0-3 edge meets the 1-2 edge in the $(x,y)$-plane, it is below in the $z$-direction. Subject to the convention \eqref{orientation} that is equivalent to the statement \eqref{lem2} about the signed volume of the tetrahedron. A purely algebraic proof can be given as follows.
\begin{proof}
Denoting the left hand side (LHS) of \eqref{orientation} by $D$, Cramer's rule gives
\begin{align}
p &= \begin{vmatrix}
x_2 - x_3 & x_2 - x_1 \\
y_2 - y_3 & y_2 - y_1
\end{vmatrix}/D, \\
q &= \begin{vmatrix}
x_0 - x_3 & x_2 - x_3 \\
y_0 - y_3 & y_2 - y_3
\end{vmatrix}/D.
\end{align}
Substituting these into \eqref{lem1} and multiplying through by $D > 0$ gives, upon expanding all the determinants, \eqref{lem2}. The reader can avoid an algebraic quagmire by referring to the Mathematica notebook provided as an ancillary file on the arXiv.
\end{proof}

If we substitute \cref{opconversion} into the CHSH inequalities we obtain \eqref{lem1} with $z_0 = c_0 x_0 + d_0 y_0 - 1$, $z_1 = -c_1 x_1 - d_1 y_1 + 1$, $z_2 = c_1 x_2 + d_1 y_2 + 1$, and $z_3 = -c_0 x_3 - d_0 y_3 - 1$, where $(c_0, d_0, c_1, d_1)$ is a column of
\begin{equation}
  \begin{pmatrix}
  1 & 1 & 1 & 1 & -1 & -1 & -1 & -1 \\
  1 & 1 & -1 & -1 & 1 & 1 & -1 & -1 \\
  1 & -1 & 1 & -1 & 1 & -1 & 1 & -1 \\
  -1 & 1 & 1 & -1 & 1 & -1 & -1 & 1
\end{pmatrix}.
\label{ineqs}
\end{equation}
(There is one column for each version of the CHSH inequality.) Substituting the $z_i$ into \eqref{lem2} then gives the desired inequality. For example the $\langle A_0B_0 \rangle + \langle A_0B_1 \rangle + \langle A_1B_0 \rangle - \langle A_1B_1\rangle \leq 2$ version of the CHSH inequality corresponds to the first column of \eqref{ineqs}, and \eqref{lem2} becomes
\begin{equation}
\begin{vmatrix}
x_0 & y_0 & x_0 + y_0 - 1 & 1 \\
x_1 & y_1 & -x_1 + y_1 + 1 & 1 \\
x_2 & y_2 & x_2 - y_2 + 1 & 1 \\
x_3 & y_3 & -x_3 - y_3 - 1 & 1
\end{vmatrix} \leq 0.
\label{standardineq}
\end{equation}
(Subject to \eqref{orientation}, this can still serve as a Bell inequality, but now in terms of $p(a|b,x,y)$ rather than $p(a,b|x,y)$.)

\section{Quantum violation}
Suppose $\Ms_0$ and $\Ms_1$ correspond to $X$ and $Z$ measurements of a qubit. Let $\Pp_0$ correspond to preparing the $+1$ eigenstate of $(X+Z)/\sqrt{2}$ and $\Pp_3$ the $-1$ eigenstate. Similarly let $\{\Pp_1, \Pp_2\}$ be the $\{+1, -1\}$ eigenstates of $(X-Z)/\sqrt{2}$. Denoting $v=1/\sqrt{2}$ the LHS of \eqref{standardineq} is
\begin{equation}
\begin{vmatrix}
v & v & 2v - 1 & 1 \\
v & -v & 1 - 2v & 1 \\
-v & v & 1 - 2v& 1 \\
-v & -v & 2v - 1 & 1
\end{vmatrix} = 16v^2(2v - 1) \approx 3.31.
\end{equation}
This is the same proof of the preparation contextuality of a qubit that appeared in \cite{parity}. However, that proof assumed $\frac12\Pp_1 + \frac12\Pp_3 = \frac12\Pp_1 + \frac12\Pp_2$, which will never hold exactly in a realistic experiment. Since no such assumption entered into \eqref{standardineq}, the proof presented here is more experimentally robust.

In \cref{quantumbell} a correspondence is established between quantum strategies in the preparation contextuality scenario and quantum strategies in the CHSH scenario. Since the above corresponds with the strategy for maximally \cite{tsirelson} violating the CHSH inequality, it is the quantum maximum for the contextuality scenario.

\section{Tomographic completeness}
A difficulty remains in the present approach, namely our assumption that $\Ms_0$ and $\Ms_1$ are tomographically complete for the $\PI$. An obvious objection in the qubit example of the previous section is the $Y$ measurement. Suppose we have a preparation noncontextual model for three measurements $\{\Ms_0, \Ms_1, \Ms'\}$, and consider a set of preparations that all give the same probability for $\Ms'$. Since \cref{opequiv} holds for all three measurements if and only if it holds for $\{\Ms_0, \Ms_1\}$, the problem reduces to finding a preparation noncontextual model for $\Ms_0$ and $\Ms_1$. Since the quantum states mentioned above all give uniformly random outcomes for the $Y$ measurement, there is in principle nothing wrong with using \eqref{standardineq}.

\begin{figure}
  \includegraphics[width=\columnwidth]{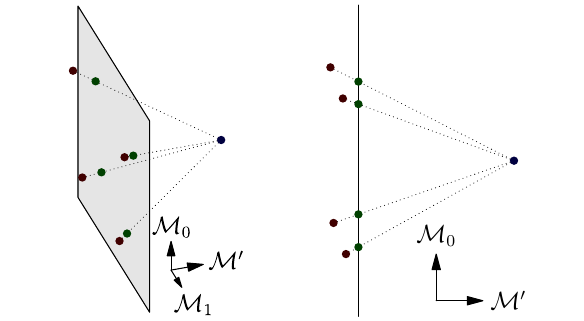}
  \caption{Dealing with failure of tomographic completeness. The statistics for three binary measurements $\{\Ms_0, \Ms_1, \Ms'\}$ are defined by a 3-dimensional vector for each preparation. Two views of this space are shown. The four non-planar preparations in brown, together with additional preparation in blue, imply the existence of the four planar preparations in green. The statistics for the green preparations can be calculated with simple trigonometry and then tested against the inequalities derived here for two binary measurements.}
  \label{tomo}
\end{figure}

In practice, however, the preparations in a real experiment will each give slightly different probabilities for the $Y$ measurement. The simplest way to deal with this is to add a fifth preparation that gives one outcome of the $Y$ measurement with high probability (e.g. the $+1$ eigenstate of $Y$). If, as in \cref{tomo}, we then consider a plane perpendicular to the $Y$ axis with the four original preparations on one side and the new preparations on the other, convexity implies the existence of, and determines the probabilities for, four preparations in the plane that can then be tested against \eqref{standardineq}. Notice there is no need to actually implement the four new preparations. Any further measurements that reveal small amounts of information about the preparations can be dealt with similarly.

This idea has been generalised \cite{exp} into a technique for identifying experimental violations of other noncontextuality inequalities, even those requiring fixed equivalences.

These techniques deal with additional measurements in the tomographically complete set, but it requires that those additional measurements are actually performed. Hence any test based on these techniques will still require the assumption that there aren't any unknown measurements that would be required to construct a tomographically complete set, i.e. unknown measurements whose statistics cannot be inferred from the measurements that have been performed. No test based on \cite{cntx} can avoid such an assumption (or some other assumption that restricts possible measurements), because it is always logically possible that there exists a measurements that simply reads out a complete description of the preparation. In that case, no two distinct preparations would be equivalent, and so any model would be trivially noncontextual.

It should be noted that an assumption about the tomographically complete set is much weaker than assuming all of quantum theory. Examples of other theories where the simplest system has three binary measurements in the tomographically complete set include Spekkens' toy theory \cite{toy} (which is noncontextual), quantum theory with fundamental decoherence \cite{dec1,dec2} (whose contextuality would depend on the amount of decoherence been preparation and measurement), and a version of Generalised No-Signalling Theory or ``boxworld'' \cite{gnst} (which could violate \cref{standardineq} \emph{more} than quantum theory). One way to lend credence to such an assumption without reference to quantum theory would be to experimentally test whether the statistics of a large number of measurements can be inferred from a small subset \cite{expgpt}. More speculatively, it may be possible to support the assumption via physical principles independent of quantum theory, for example thermodynamical principles (c.f. \cite{thermo}).

\section{Conclusions} The eight inequalities derived here fully classify the preparation noncontextual statistics in the simplest non-trivial scenario. No assumptions of unattainable operational equivalences were made.  No new assumptions on the representation of approximately operationally equivalent procedures \cite{mod1,mod2} were made either. Furthermore, it was not assumed that the measurements are represented deterministically in the ontological model, which (in quantum terms) means it doesn't matter whether the measurements are projective \cite{cntx,robrant}. Hence a violation can only be explained by a failure of noncontextuality or a failure of the tomographic completeness of the measurements. Since observed failures of the latter can be dealt with using convexity, nonclassicality may be left as the only plausible explanation.

The main extension of these results would be to classify scenarios with more preparations and measurements. Additional (performed) measurements in the tomographically complete set could then be dealt with more elegantly than above, by fully incorporating the extra procedures into the contextuality scenario. In \cref{gensec} some of the results above are generalised to such scenarios, and the limitations of these generalisations are discussed. It would also be interesting to apply similar ideas to measurement and transformation noncontextuality. More broadly, the status of tomographic completeness assumptions in tests of contextuality deserves further study.

\begin{acknowledgments}
Thanks to Rob Spekkens for invaluable discussions, to L\'idia del Rio for feedback on a previous version, and to Jon Barrett for sharing his result \cite{jon} and allowing me to include it here. Research at Perimeter Institute is supported in part by the Government of Canada through NSERC and by the Province of Ontario through MRI. I am now supported by the Royal Commision for the Exhibition of 1851.
\end{acknowledgments}

\bibliography{ctx}
\appendix
\section{Formulating preparation noncontextuality}\label{formulating}

In \cite{cntx}, preparation noncontextuality is defined as the requirement that if $P(k|\Pp,\Ms) = P(k|\Pp',\Ms)$ for all $\Ms$, then $P(\lambda|\Pp) = P(\lambda|\Pp')$. Here I show if we are only interested in a finite set of preparations $\{\PI\}$ (and convex combinations thereof), and we assume the $\MJ$ are tomographically complete, then this definition is equivalent to the definition in \cref{definitions}, i.e. the requirement that if \cref{opequiv} holds then \cref{onticequiv} holds.

Define $\Pp$ as the procedure of preparing $\PI$ with probability $p_i$, and $\Pp'$ as the procedure of preparing $\PI$ with probability $q_i$, and notice that any of the preparations we are interested in are of this form (in particular $\{p_i\}$ can assign probability 1 to a preparation.) By definition $P(k|\Pp,\Ms) = \sum_i p_i P(k|\PI,\Ms)$ and $P(k|\Pp',\Ms) = \sum_i q_i P(k|\PI,\Ms)$. Hence \cref{opequiv} is exactly the statement that $P(k|\Pp,\MJ) = P(k|\Pp',\MJ)$ for all $j$. Tomographic completeness means that this is equivalent to $P(k|\Pp, \Ms) = P(k|\Pp', \Ms)$ for all $\Ms$. So the ``if'' conditions of the two definitions are equivalent.

Again by the definitions of $\Pp$ and $\Pp'$, $P(\lambda|\Pp) = \sum_i p_i\mu_i(\lambda)$ and $P(\lambda|\Pp') = \sum_i q_i \mu_i(\lambda)$ and so \cref{onticequiv} is exactly $P(\lambda|\Pp) = P(\lambda|\Pp')$. So the ``then'' implications of each definition are also equivalent, and we are done.

\section{Identifying the simplest scenario}\label{simplest}
Here I consider scenarios simpler than the four preparations and two binary measurements discussed in the main text, and show that they all trivially admit noncontextual models for any values of the operational probabilities $P(k|\PI,\MJ)$. The basic ideas are as follows. For one measurement, a noncontextual model can be obtained by having the ontic state encode the outcome. On the other hand, if we have so few preparations that they form a simplex in the space of operational probabilities then the ontic state can simply encode the identity of the preparation.

In detail, suppose we consider only a single measurement $\Ms_1$. There is a simple ``$\lambda = k$'' model, where the ontic state simply specifies the outcome of $\Ms_1$, i.e. $P(k | \lambda, \Ms_1) = \delta_{k\lambda}$. \Cref{onticop} is ensured by distributing the ontic states according to $\mu_i(\lambda) = P(k = \lambda | \PI, M_1)$. This is manifestly preparation noncontextual because it makes \cref{onticequiv} the same as \cref{opequiv}. Hence the simplest non-trivial scenario must have at least two measurements, the simplest such case being two binary (two-outcome) measurements.

Now suppose we consider three or fewer preparations with two binary measurements. As noted in \cref{definitions}, we can associate preparations in such a scenario with vectors $\vec \PI = (P(0|\PI,\Ms_0)-P(1|\PI,\Ms_0), P(0|\PI,\Ms_1)-P(1|\PI,\Ms_1))$. The convex hull of any one to three $\vec \PI$ is a simplex: a point, a line segment, or a triangle. Recall that every point in a simplex has exactly one decomposition into extremal points. Hence we consider the ontological model in which there is an ontic state $\lambda$ for each extreme point $\vec\lambda$ of that simplex, with $\mu_i(\lambda)$ being the unique distribution ensuring that $\sum_\lambda \mu_i(\lambda)\vec \lambda = \vec\PI$. Naturally $P(k|\lambda,\Ms_i)$ is defined so that $\vec \lambda = (P(0|\lambda,\Ms_0)-P(1|\lambda,\Ms_0), P(0|\lambda,\Ms_1)-P(1|\lambda,\Ms_1))$, so that \cref{onticop} is satisfied. Expanding \cref{opequiv} as
\begin{equation}
  \sum_{i,\lambda} p_i \mu_i(\lambda) \vec \lambda = \sum_{i,\lambda} q_i \mu_i(\lambda) \vec\lambda
\end{equation}
and applying uniqueness again immediately gives \cref{onticequiv}, so our model is preparation noncontextual.

\section{Quantum connection with the Bell scenario}\label{quantumbell}
In \cref{chshsec} it is shown that for the scenario considered, the $P(k|\PI,\MJ)$ admit a preparation noncontextual model if and only if the $P(a,b|x,y)$ defined in \cref{opconversion} admit a locally causal model. Here I sketch an argument that, similarly, the $P(k|\PI,\MJ)$ admit a quantum realisation (satisfying the tomographic completeness assumption) if and only if the $P(a,b|x,y)$ are quantum-realisable in the usual sense that
\begin{equation}
  P(a,b|x,y) = \tr\left((E_{a|x}\otimes F_{b|y})\rho_{AB}\right)
\end{equation}
for a bipartite quantum state $\rho_{AB}$ and sets of POVMs $\{E_{a|x}\}$ and $\{E_{b|y}\}$.

For the ``if part'', let $\rho_i \propto \tr_A\left( (E_i \otimes I) \rho_{AB} \right)$ be the normalised steered state when Alice measures $E_0 = E_{0|0}, E_1 = E_{0|1}, E_2 = E_{1|1}, E_3 = E_{1|0}$ respectively. The basic idea is that in the noncontextuality scenario $\PI$ will correspond to preparing $\rho_i$ and $\MJ$ to measuring $\{F_{k|j}\}$. Certainly this will achieve the correct $P(k|\PI,\MJ)$ according to \cref{opconversion}. However we also need that the $\MJ$ are tomographically complete for the $\PI$. Plotting the $P(k|\PI,\MJ)$ as in \cref{conventions}, there are two cases to consider. The first is that the $\PI$ form a two-dimensional shape. In this case the $\MJ$ must be tomographically complete because the $\rho_i$ certainly live in \emph{some} two-dimensional affine subspace of the quantum states by the no-signalling condition $p\rho_0 + (1-p)\rho_3 = q\rho_1 + (1-q)\rho_2 = \tr_A(\rho_{AB})$, and so the only way for the projection onto the $\MJ$ to be two-dimensional is if they span that space. The second case is that the $\PI$ form a one-dimensional shape, which is necessarily a simplex. Hence the statistics can be reproduced using the preparation noncontextual ``extreme point'' model of the previous section, which can be implemented with the ontic state encoded in a qubit and the $\MJ$ will be tomographically complete by construction.

For the ``only if'' part, there are again two cases that arise on consideration of the geometry in \cref{conventions}. The first case is that the $\PI$ indeed form a non-degenerate convex quadrilateral as shown in the figure. Adopting the specified labelling convention, we then have $p\rho_0 + (1-p)\rho_3 = q\rho_1 + (1-q)\rho_2 =: \rho_B$ for some probabilities $p$ and $q$ (we are using tomographic completeness to derive that the two mixtures having the same statistics for the $\MJ$ implies they correspond to the same density operator). Let $\rho_{AB}$ be a purification of $\rho_{B}$. By the Schr\"odinger-HJW theorem \cite{schr,HJW} there exist measurements for Alice that steer Bob onto the two decompositions of $\rho_B$, giving the required measurements for the correct $P(a,b|x,y)$ in the Bell scenario. The second case is that the convex hull of the $\PI$ is a simplex, in which case there is a noncontextual model by the argument in the previous section, and hence a Bell-local model for the corresponding $P(a,b|x,y)$ by the argument in \cref{chshsec}. Any Bell-local $P(a,b|x,y)$ is also a quantum $P(a,b|x,y)$.

\section{Generalisations to other scenarios}\label{gensec}
Here I provide generalisations of some of the steps in the main text to scenarios involving more preparations or measurements, and discuss the difficulties in using these techniques to provide a full classification of such scenarios.

Notice that if any finite number of measurements are tomographically complete, the preparations $\PI$ are characterised by a finite number of probabilities and hence can be considered as points in a finite-dimensional vector space. This generalises the two-dimensional space of $\vec \PI$ discussed in the main text, for notational simplicity we will not distinguish a preparation from its vector here.

\subsection{Only need to look at decompositions of one state}
The following is a straightforward generalisation to arbitrary numbers of preparations and measurements of the first step in \cref{pivotalsec}.
\begin{lem}
  Let $\Pp^*$ be in the interior of the convex hull of the $\{\PI\}$. An ontological model for the $\{\PI\}$ is preparation noncontextual if and only if there exists a $\mu^*(\lambda)$ such that for all probability distributions $\{p_i\}$ such that
\begin{equation} \sum_i p_i \PI = \Pp^*, \label{staropequiv}
\end{equation}
we have
\begin{equation}
  \sum_i p_i \mu_i(\lambda) = \mu^*(\lambda).\label{staronticequiv}
\end{equation}
\label{starlem}
\end{lem}

\begin{proof} The ``only if'' part is trivial, simply take $\mu(\lambda) = \sum_i p_i \mu_i(\lambda)$ for one decomposition of $\Pp^*$ and note that preparation noncontextuality ensures this works for any other decompositions.

For the ``if'' part, suppose that an ontological model is preparation contextual. In particular, there exists distributions $\{p_i\}$ and $\{q_i\}$ such that \cref{opequiv} holds and yet \cref{onticequiv} fails. Define $\tilde \Pp = \sum_i p_i \PI$ ($=\sum_i q_i \PI$). Since $\Pp^*$ is in the interior of the state space, $\Pp^* = \sum_i r_i \PI + \tilde r \tilde\Pp$ for some probability distribution $\{r_i, \tilde r\}$ with $\tilde r > 0$. But then letting $p_i' = r_i + \tilde r p_i$ and $q_i' = r_i + \tilde r q_i$ give two distributions that satisfy \cref{staropequiv}, yet the failure of \cref{onticequiv} for $p_i, q_i$ and $r_* > 0$ means that $\sum_i p_i' \mu_i(\lambda) \neq q_i'\mu_i(\lambda)$. Hence \cref{staronticequiv} cannot be satisfied for both $\{p_i'\}$ and $\{q_i'\}$.
\end{proof}

\subsection{Only need to look at a finite number of decompositions}
Now a somewhat less straightforward generalisation of the second step in \cref{pivotalsec}.

\begin{thm}
Let $\Pp^*$ be in the interior of the state space. An ontological model is preparation noncontextual if and only if there exists a $\mu^*(\lambda)$ such that for all probability distributions $\{p_i\}$, with $\{\PI : p_i > 0\}$ forming a simplex, and such that
\begin{equation} \sum_i p_i \PI = \Pp^*, \label{finopequiv}
\end{equation}
we have
\begin{equation}
  \sum_i p_i \mu_i(\lambda) = \mu^*(\lambda).\label{finonticequiv}
\end{equation}
Furthermore, there are a finite number of such $\{p_i\}$.
\label{finthm}
\end{thm}

We will need the following:
\begin{lem}
Consider a real $d$-dimensional vector space $V$, a finite set of points $v_i \in V$, and a further point $v \in V$. The set of probability distributions $\{ p_i \}$ such that $\sum_i p_i v_i = v$ is a closed convex polytope, the extreme points of which are exactly those elements supported on $i$ such that the $v_i$ form a simplex.
\label{finlem}
\end{lem}

\begin{proof}
Such $p_i$ are defined by the finite set of linear constraints $p_i \geq 0$, $\sum_i p_i = 1$ and $\sum_i p_i v_i = v$, hence forming a closed convex polytope.

Suppose $\{ p^*_i \}$ is an element of the polytope, and that its support $S^* = \{ i : p^*_i > 0 \}$ defines a non-simplex $v_i$ for $i \in S^*$. Hence the size of $S^*$ must be at least $d^* + 2$, where $d^*$ is the dimension of the affine span of the $v_i$ with $i \in S$. By Carath\'{e}odory's theorem in convex geometry there exists another element $\{p'_i \}$ of the polytope, supported on a proper subset of $S^*$. But then for a sufficiently small value of $q$, with $0 < q < 1$, $p''_i = (p^*_i - qp'_i)/(1-q)$ will satisfy $0 \leq p''_i \leq 1$. Noting that $p''_i$ is then another element of the polytope, and that $p^*_i = qp'_i + (1-q)p''_i$, we see that $\{p^*_i\}$ is not extremal.

On the other hand, suppose $\{ p^*_i\}$ has support $S^* = \{ i : p^*_i > 0 \}$ such that the $v_i$ form a simplex. Any other elements of the polytope that $\{ p^*_i\}$ can be decomposed into must have the same (or smaller) support. But there is only one way to write an element of a simplex as a convex combination of the vertices of the simplex, and so the decomposition must be trivial. That is to say, $\{ p^*_i\}$ is extremal.
\end{proof}

\begin{proof}[Proof of \cref{finthm}]
The ``only if'' part is again trivial.

For the ``if'' part, by \cref{starlem} we only need to ensure that $\mu^*$ satisfies \cref{staronticequiv} whenever $\{p_i\}$ satisfies \cref{staropequiv}. Letting $v_i$ and $v$ represent $\PI$ and $\Pp^*$ respectively in \cref{finlem}, we see that any $\{ p_i\}$ satisfying \cref{staropequiv} is an element of the described polytope. Any element of a polytope can be written as a convex combination of its extreme points, and so there exists a distribution $q_j$ such that
\begin{equation} \sum_j q_j p^{(j)}_i = p_i, \label{fineq1}
\end{equation}
where for each $j$, $\{p^{(j)}_i\}$ is a distribution supported on $i$ such that $\PI$ form a simplex. But then by assumption we have
\begin{equation} \sum_i p^{(j)}_i \mu_i(\lambda) = \mu^*(\lambda). \label{fineq2}
\end{equation}
Combining \cref{fineq1} with \cref{fineq2} we find
\begin{equation} \sum_i p_i \mu_i(\lambda) = \sum_{i,j} q_j p^{(j)}_i \mu_i(\lambda) = \sum_j q_j \mu^*(\lambda) = \mu^*(\lambda) \end{equation}
as required.
\end{proof}

In the main text the special case of four preparations forming a quadrilateral with $\Pp^*$ at the intersection of the diagonals was used. In that case \cref{finthm} shows that you only need to check the two decompositions onto opposite pairs of corners.

\subsection{The connection with Bell's theorem}
In \cref{chshsec} the preparation contextuality scenario under consideration was linked with the CHSH scenario based on the ideas of \cite{jon}. In light of the above results, one might hope that more complicated preparation contextuality scenarios can be also be fully characterised by translating to Bell scenarios. Unfortunately that appears not to be the case: even though \cref{finthm} reduces the problem to a finite number of decompositions of a single preparation, the same $\PI$ will normally appear in multiple decompositions. As far as I can see there is no way to enforce that the corresponding $\mu_i$ are the same when converting to a Bell scenario without breaking the linearity that is so important computationally. This leaves the cases where the sets of $\PI$ happen to be disjoint:

\begin{thm}
Suppose there exists a $\Pp^*$ such that each $i$ appears in the support of exactly one of the decompositions $\{p_i^{(1)}\}$, $\{p_i^{(2)}\}$, \dots, described in \cref{finthm}. Then there exists a preparation noncontextual model if and only if the bipartite probabilities
\begin{equation}
P(i,k|x,j) = p_i^{(x)} P(k|\PI, \MJ)
\label{bellprobs}
\end{equation}
admit a locally causal model.
\label{bellthm}
\end{thm}

\begin{proof}
For the ``if'' part: starting with a locally causal model
\begin{equation}
  P(i,k|x,j) = \int P(i|\lambda,x) P(k|\lambda,\MJ) \mu^*(\lambda) d\lambda,
  \label{belllocal}
\end{equation}
we can define an ontological model for the contextuality scenario by
\begin{equation}
  \mu_i(\lambda) = \frac{\mu^*(\lambda)P(i|\lambda, x(i))}{p_i^{(x(i))}}
  \label{bellalice}
\end{equation}
where $x(i)$ is uniquely defined by $p^{(x(i))}_i > 0$.

By construction this model satisfies the requirement of \cref{finthm} and so is preparation noncontextual. By \cref{bellprobs,belllocal,bellalice} its operational predictions are
\begin{widetext}
\begin{equation}
  \int P(k|\lambda,\MJ) \mu_i(\lambda)d\lambda = \int P(k|\lambda,\MJ)\frac{\mu^*(\lambda)P(i|\lambda, x(i))}{p_i^{(x(i))}} d\lambda
  = \frac{P(i,k|x(i), j)}{p_i^{(x(i))}} = P(k|\PI,\MJ)
\end{equation}
as required.

For the ``only if'' part: starting with a preparation noncontextual model $\mu_i(\lambda)$ and $P(k|\lambda, \MJ)$ we can define a locally causal model using the $\mu^*$ from \cref{finthm} and
\begin{equation}
  P(i|\lambda,x) = \frac{\mu_i(\lambda)p_i^{(x)}}{\mu^*(\lambda)},
\end{equation}
which gives
\begin{equation}
  \int P(i|\lambda,x) P(k|\lambda,\MJ) \mu^*(\lambda) d\lambda
= \int \mu_i(\lambda)p_i^{(x)}P(k|\lambda,\MJ)d\lambda = p_i^{(x)}P(k|\PI,\MJ) = P(i,k|x,j)
\end{equation}
\end{widetext}
as required.
\end{proof}
The ``only if'' part does not require the assumption that each $i$ appears in only one support. Hence, in any contextuality scenario, one way to show the impossibility of a preparation noncontextual model is to show that \cref{bellprobs} violate a Bell inequality. The special thing about scenarios where each $i$ appears in only one support is that this technique is powerful enough to capture \emph{every} failure of preparation noncontextuality.

A generalisation of \cref{closedsec} is left open.
\end{document}